\documentclass[aps,prd,superscriptaddress,nofootinbib,eqsecnum]{revtex4}

\pdfoutput=1

\usepackage{amsfonts}
\usepackage{amsmath}
\usepackage{amssymb}
\usepackage{graphicx,color}
\usepackage{float}
\usepackage{hyperref}
\usepackage{subfigure}
\usepackage{mathtools,slashed}

\begin{document}
\title{Effect of the magnetized medium on the decay of neutral scalar bosons}

\author{Aritra Bandyopadhyay}
\affiliation{Departamento de Fisica, Universidade Federal de Santa Maria, Santa Maria, RS 97105-900, Brazil}

\author{Ricardo L. S. Farias}
\affiliation{Departamento de Fisica, Universidade Federal de Santa Maria, Santa Maria, RS 97105-900, Brazil}

\author{Rudnei O. Ramos}
\affiliation{Departamento de F\'{\i}sica Te\'orica, Universidade do
  Estado do Rio de Janeiro, 20550-013 Rio de Janeiro, RJ, Brazil}

\begin{abstract}

The decay of a heavy neutral scalar particle into fermions and into
charged scalars are analyzed when in the presence of an external
magnetic field and finite temperature.  Working in the one-loop
approximation for the study of these decay channels, it is shown that
the magnetic field leads in general to a suppression of the decay width
whenever the kinematic constrain depends explicitly on the magnetic field.
Our results are also compared with common approximations found in
the literature, e.g., when the magnitude of the external magnetic
field is smaller than the decaying product particle masses, i.e., in
the weak field approximation, and in the opposite case, i.e., in the
strong field approximation.  Possible applications of our results are
discussed.
 
\end{abstract}

\maketitle

\section{Introduction}

Magnetic fields are omnipresent in the Universe, where we can find
fields with magnitude ranging from as low as around $10^{-16}$ Gauss
in the intergalatic medium~\cite{Subramanian:2015lua}, to up to around
$10^{15}$ Gauss  in strongly magnetized neutron stars (or
magnetars)~\cite{Kaspi:2017fwg}.  Magnetic fields of even larger
magnitude can also be found in terrestrial laboratory experiments.
{}For instance, at the Relativistic Heavy Ion Collider (RHIC) and the
Large Hadron Collider (LHC) facilities   magnetic fields as large as
$10^{18}$ to $10^{19}$ Gauss can be
produced~\cite{Skokov:2009qp,Deng:2012pc}.  Cosmological phase
transitions that might have happened in the early Universe are another
potential source of generation of strong magnetic fields. {}For
instance, at the electroweak phase transitions it is supposed that
magnetic fields with strength of order of $10^{23}$ Gauss could be
produced~\cite{Vachaspati:1991nm}.

The presence of magnetic fields have the ability of influencing many
physical processes over a broad range of scales in the Universe.
Their effects can be important already at the time they are formed
during the very early cosmological phase
transitions~\cite{Linde:1978px}. It is also well-known that the
presence of magnetic fields at the time of recombination and the
cosmic microwave background (CMB)  radiation formation can lead to
anisotropies in the CMB~\cite{Grasso:2000wj,Widrow:2002ud,Giovannini:2017rbc}. It can
also affect the big-bang nucleosynthesis  epoch changing the light
nuclei formation, affect the formation of the early stars,  among
other important
consequences~\cite{Giovannini:2003yn,Widrow:2011hs}. All these effects
can severely constrain the magnitude of the magnetic field present in
the Universe at those early times.

Particular emphasis has been given also to the effects of the high
magnitude magnetic fields  generated in the heavy ion collision
experiments mentioned above.  {}For instance, these experiments have
given enough indications for the formation of a deconfined state of
hadronic matter, called quark gluon plasma (QGP) under extreme
conditions of high densities and temperatures (see, e.g.,
Refs.~\cite{Shuryak:2014zxa,Pasechnik:2016wkt} for reviews). Recently
a captivating nature of non-central heavy ion  collisions has come
into light, the generation of a rapidly decaying strong anisotropic
magnetic field in the  direction perpendicular to the reaction plane,
due to the relative motion of the ions themselves.  The nature of the
decay in the magnitude of this magnetic field has been a subject of
debate as  some of the studies reveal that it decreases very fast,
being inversely proportional to the square of
time~\cite{Bzdak:2012fr,McLerran:2013hla}, whereas  other studies opt
for an adiabatic decay due to high conductivity of the
medium~\cite{Tuchin:2013bda,Tuchin:2012mf,Tuchin:2013ie}.  These
findings have also sparkled an intense research activity to study the
properties of strongly  interacting matter in presence of an external
magnetic field resulting in the emergence of several novel phenomena,
e.g., the  finite temperature magnetic
catalysis~\cite{Shovkovy:2012zn,Alexandre:2000yf,Gusynin:1997kj,Lee:1997zj}
and the inverse  magnetic
catalysis~\cite{Bali:2011qj,Bruckmann:2013oba,Mueller:2015fka,Ayala:2014iba,Ayala:2014gwa,Ayala:2015bgv,Farias:2014eca,Farias:2016gmy}
as some of the examples of these effects. 

The possible consequences caused by magnetic fields in different
systems in nature demonstrate that  there is clearly an increasing
demand to understand their role in many physical phenomena.  In the
present work, we will be particularly concerned in understanding the
effects of intense  external background magnetic fields on particle
decay processes. Some previous studies of decay processes  in a
magnetized medium include for example the ones done in the
Refs.~\cite{Bali:2018sey,Bandyopadhyay:2016cpf,Tsai:1974fa,Kuznetsov:1997iy,Piccinelli:2017yvl,Kawaguchi:2016gbf,Satunin:2013an,Sogut:2017ksu,Ghosh:2017rjo}.
The different methods and approximations used in those previous
literature have lead to some conflicting results for the behavior of
decay rates as a function of the background external magnetic field.
{}For example, while some studies show that the decay widths can be
enhanced through the effect of magnetic
medium~\cite{Bali:2018sey,Bandyopadhyay:2016cpf,Tsai:1974fa,Kuznetsov:1997iy,Satunin:2013an},
others show a suppression
effect~\cite{Piccinelli:2017yvl,Sogut:2017ksu,Kawaguchi:2016gbf}. Even
a  mixed behavior for different energies is also found in
Ref.~\cite{Ghosh:2017rjo}. In our present work, we analyze in details
the case of the decay of a neutral scalar bosons into a fermion and an
antifermion and also the case of decay into charged scalar
particles. We study different limiting cases, as well as the
most general scenario with arbitrary magnitude of the  external
magnetic field to gauge the validity of each of these
approximations.  This way, one should be able to understand the possible sources of the differences found in the literature.

This work is organized as follows.  In Sec.~\ref{sec2} we give the
relevant definitions and equations needed to derive the decay width
for a real scalar field with an Yukawa interaction to fermions and in
the presence of a magnetic field and finite temperature.  The decay
width is then explicitly derived.  In Sec.~\ref{limitingcases} we
study the two limiting cases for the decay into fermions, namely, the
weak magnetic field and the strong magnetic field approximations. In Sec.~\ref{results} we discuss the different results
obtained for the decay width into a pair of fermion and antifermion.
In Sec.~\ref{boson} we turn our attention to the similar study of the
decay of a heavy neutral scalar field into charged scalars. Our
concluding remarks along with a discussion of possible applications of our
results are given in Sec.~\ref{conclusions}.  Two Appendices are
included  where we give some of the technical details of the relevant
calculations.

\section{Fermionic decay in the presence of a constant magnetic field}
\label{sec2}

The primary ingredient of the theoretical tools for studying the
various decay processes in quantum field theory is the $n$-point
correlation function. By the virtue of the optical
theorem~\cite{Peskin:1995ev} one can connect the  imaginary part, or
the discontinuity, of the two-point correlation function, e.g., for a
scalar particle,  $\mathrm{Im}~\Pi$, with the decay width $\Gamma$ of
an unstable particle in the rest frame of the decaying scalar via the
relation
\begin{eqnarray}
\Gamma = \frac{\mathrm{Im}~\Pi(P)}{M},
\end{eqnarray}
where $M$ is the invariant mass of the decaying scalar, which is
equivalent to the four-momentum $P$ of the same.  Hence, let us
initially focus our study in the one-loop (leading order) self-energy
function of a  neutral heavy boson decaying into two light fermions
and when in the presence of an external magnetic field. 

\subsection{Dirac propagator in an external magnetic field}

In the following, we will make use of the Schwinger's proper time
propagator~\cite{Schwinger:1951nm,Schwinger:book}.  The charged
fermion propagator in  coordinate space is then expressed as

\begin{eqnarray}
S_m(x,x^\prime) =
e^{\Phi(x,x^\prime)}\int\frac{d^4K}{(2\pi)^4}e^{-iK(x-x^\prime)}S_m(K)
, 
\end{eqnarray}
where $\Phi(x,x^\prime)$ is called the phase
factor~\cite{Schwinger:1951nm,Schwinger:book}, which generally drops
out in gauge  invariant correlation functions and the exact form of
$\Phi(x,x^\prime)$ is not important in our problem when evaluating the
fermion loop self-energy relevant for the determination of the decay
width. In momentum space the  Schwinger propagator $S_m(K)$ is written
as an  integral over proper time $s$, 
\begin{eqnarray}
i S_m(K) = \int\limits_0^\infty ds \exp\left[i s
  \left(K_\shortparallel^2-m_f^2-\frac{K_\perp^2}{q_f Bs}\tan(q_f
  Bs)\right)\right] \nonumber\\ \times
\left[\left(\slashed{K}_\shortparallel+m_f\right)\left(1+\gamma_1\gamma_2\tan(q_f
  B s)\right)-\slashed{K}_\perp\left(1+\tan^2(q_f Bs)\right)\right].
\label{schwinger_propertime}
\end{eqnarray}
where we are considering the case of a constant magnetic field
pointing towards the $z$ direction, $\vec{B}=B\hat{z}$.  Here, $m_f$
and $q_f$ are the mass and absolute charge of the fermion,
respectively, whereas $\shortparallel$ and $\perp$ are, respectively,
the parallel and perpendicular  components of the momentum, which are
now separated out in the momentum space propagator. 

We will follow the notation where
\begin{eqnarray}
&&a^\mu = a_\shortparallel^\mu + a_\perp^\mu, \nonumber\\ &&
  a_\shortparallel^\mu = (a^0,0,0,a^3), \nonumber\\   &&a_\perp^\mu =
  (0,a^1,a^2,0), \nonumber
\end{eqnarray}
with the metric signature defined as
\begin{eqnarray}
&&g^{\mu\nu} = g_\shortparallel^{\mu\nu} + g_\perp^{\mu\nu}, \nonumber
  \\  &&g_\shortparallel^{\mu\nu}= \textsf{diag}(1,0,0,-1), \nonumber
  \\  &&g_\perp^{\mu\nu} = \textsf{diag}(0,-1,-1,0), \nonumber
\end{eqnarray}
such that
\begin{eqnarray}
&&(a\cdot b) = (a\cdot b)_\shortparallel - (a\cdot b)_\perp, \nonumber
  \\  &&(a\cdot b)_\shortparallel = a^0b^0-a^3b^3,  \nonumber
  \\ &&(a\cdot b)_\perp = (a^1b^1+a^2b^2).\nonumber
\end{eqnarray}

Using the identity
\begin{eqnarray}
i\tan(x) = \frac{1-\exp(-2ix)}{1+\exp(-2ix)},
\end{eqnarray}
the proper time integration in Eq.~(\ref{schwinger_propertime}) can be
performed and  the fermion propagator can then be represented as a sum
over discrete energy spectrum for the fermion~\cite{Chodos:1990vv,Gusynin:1995nb,Mukherjee:2017dls}, 
\begin{eqnarray}
i S_m(K) = i e^{-\frac{K_\perp^2}{q_fB}} \sum_{n=0}^{\infty}
\frac{(-1)^nD_n(q_fB, K)}{K_\shortparallel^2-m_f^2-2nq_fB},
\label{decomposed_propagator}
\end{eqnarray}
with $n=0,\, 1,\, 2, \ldots$, denoting the Landau levels  and  
\begin{eqnarray}
D_n(q_fB,K) &=&
(\slashed{K}_\shortparallel+m_f)\Bigl((1-i\gamma^1\gamma^2)L_n\left(\frac{2K_\perp^2}{q_fB}\right)
\nonumber\\ &-&(1+i\gamma^1\gamma^2)L_{n-1}\left(\frac{2K_\perp^2}{q_fB}\right)\Bigr)-
4\slashed{K}_\perp L_{n-1}^1\left(\frac{2K_\perp^2}{q_fB}\right),
\label{d_n}
\end{eqnarray}
where $L_n^\alpha (x)$ is the generalized Laguerre polynomial, defined
as
\begin{eqnarray}
(1-z)^{-(\alpha+1)}\exp\left(\frac{xz}{z-1}\right) =
  \sum_{n=0}^{\infty} L_n^\alpha(x) z^n
\end{eqnarray}
and satisfying the property $L_{-1}^\alpha(x)=0$.

\subsection{The one-loop scalar field self-energy and its imaginary part}

We consider a real scalar field $\Phi$ interacting with the fermion
field $\psi$, through the Yukawa interaction, 
\begin{eqnarray}
\mathcal{L_{\rm int}} = g\Phi\bar{\psi}\psi.
\label{Lint}
\end{eqnarray}

\begin{center}
\begin{figure}[!htb]
\includegraphics[width=6cm]{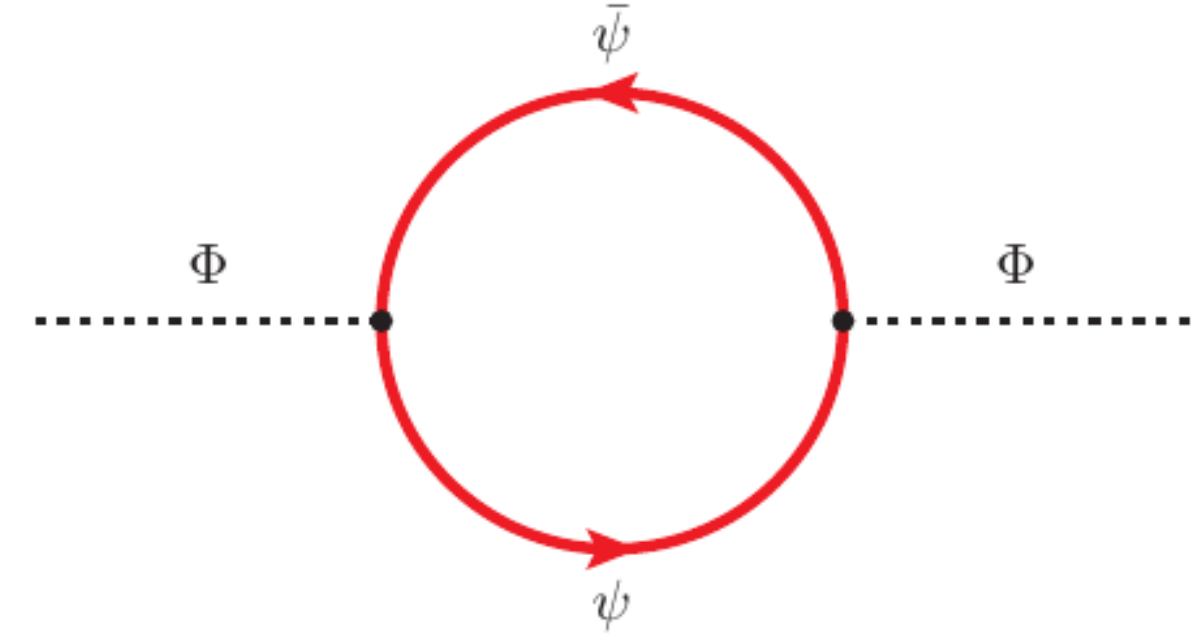}
\caption{The one-loop self-energy of a neutral scalar $\Phi$ boson
  with an Yukawa interaction with a fermion field.}
\label{fig1}
\end{figure}
\end{center}

Given the interacting Lagrangian (\ref{Lint}), the one-loop
self-energy for $\Phi$ is given by  the {}Feynman diagram shown in
{}Fig.~\ref{fig1}.  Its explicit expression is given by 
\begin{eqnarray}
i\Pi(P) &=& (ig)^2 \int\frac{d^4K}{(2\pi)^4}
\mathrm{Tr}\left[S_{m}(-Q)S_{m}(K)\right],
\end{eqnarray}
where we have denoted $Q=P-K$ and $P$ is the external four-momentum.

Using the expression for the fermion propagator decomposed in terms of
different Landau levels  in presence of any arbitrary magnetic field
given by Eq.~(\ref{decomposed_propagator}), the self-energy becomes
\begin{eqnarray}
i\Pi(P)&=& -g^2 \int\frac{d^4K}{(2\pi)^4}~
e^{-\frac{Q_\perp^2+K_\perp^2}{q_fB}}\sum_{n=0}^{+\infty}\sum_{m=0}^{+\infty}
(-1)^{m+n}\left(\frac{\textsf{Tr}\left[
    D_n(q_fB,-Q)~D_m(q_fB,K)\right]}{(Q_\shortparallel^2-m_f^2-2nq_fB)(K_\shortparallel^2-m_f^2-2mq_fB)}
\right),
\label{arbit_self_initial_expression}
\end{eqnarray}
where the expressions for $D_n$ is given by Eq.~(\ref{d_n}).  The
details of the derivation of Eq.~(\ref{arbit_self_initial_expression})
is given in the Appendix~\ref{appA}.  The final result can be
expressed as 
\begin{eqnarray}
i\Pi(P_\shortparallel) &=& \frac{g^2q_f B}{\pi} \sum_{n=0}^{+\infty}
(2-\delta_{n,0})
\int\frac{d^2K_\shortparallel}{(2\pi)^2}~\frac{(K\cdot
  Q)_\shortparallel-m_f^2+4 n q_f B} {(Q_\shortparallel^2-m_f^2-2n q_f
  B)(K_\shortparallel^2-m_f^2-2n q_f B)},
\label{fullPi}
\end{eqnarray}
where we have set $P_\perp =0$ (we will mainly be interested in the
expression for the decay width in the rest frame of the decaying
scalar particle). 

The imaginary part of the self-energy determines the decay width of
the heavy boson, which in the onshell and rest frame of the decaying
particle is defined as~\cite{Peskin:1995ev}
\begin{eqnarray}
\Gamma_M = \frac{\mathrm{Im}~\Pi(P=M)}{M}.
\label{decay}
\end{eqnarray}
Hence, by evaluating the imaginary part of Eq.~(\ref{fullPi}) at
finite temperature and in the presence of the external magnetic field,
we obtain (see Appendix~\ref{appA})
\begin{eqnarray}
\Gamma_M(B,T) &=& \frac{g^2q_fB}{2 \pi M}\sum_{n=0}^{+\infty} (2 -
\delta_{n,0})\left(1-\frac{4m_f'^2}{M^2}
\right)^{{1}/{2}}\Theta\left(M^2-4m_f'^2\right)\Bigl[1-2n_F\left(\frac{M}{2}\right)\Bigr]\nonumber\\ &&+\frac{2g^2(q_fB)^2}{\pi
  M^3}\sum_{n=1}^{+\infty} n\left(1-\frac{4m_f'^2}{M^2}
\right)^{-{1}/{2}}\Theta\left(M^2-4m_f'^2\right)\Bigl[1-2n_F\left(\frac{M}{2}\right)\Bigr],
\label{fulldecay}
\end{eqnarray}
where we have defined $m_f'^2= m_f^2 + 2 n q_f B$. The expression
(\ref{fulldecay}) gives the impression  that it would diverge as we
approach the threshold from above, $M\to 2m_f'^+$, but this is
misleading.  In fact, one notes that the kinematic constrain, set by
the Heaviside function $\Theta\left(M^2-4m_f'^2\right)$, implies that
the sum is constrained up to a maximum value integer value $N_{\rm
  max}(B)$,  given in terms of the magnetic field as
\begin{equation}
n < N_{\rm max}(B) = {\rm Integer}\left[\frac{M^2-4m_f^2}{8 q_f
    B}\right]. 
\label{condition}
\end{equation}
Explicitly, Eq.~(\ref{fulldecay}) then becomes
\begin{eqnarray}
\Gamma_M(B,T) &=& \frac{\sqrt{2}g^2\left(q_fB\right)^{3/2}}{\pi M^2}
\left[
\sqrt{\frac{M^2-4m_f^2}{8 q_f B}} \; \Theta\left(M^2-4m_f^2\right)
 \right.  \nonumber \\ &+&
  \left. 
\sum_{n=1}^{N_{\rm max}(B)-1} \left(2 
  \sqrt{\frac{M^2-4m_f^2}{8 q_f B}-n} + \frac{1}{2} 
  \frac{n}{\sqrt{\frac{M^2-4m_f^2}{8 q_f B}-n}} \right)\right]
\left[1-2n_F\left(\frac{M}{2}\right)\right],
\label{fulldecaysum}
\end{eqnarray}
where we have explicitly separated the LLL ($n=0$) term in the above expression.
Note that for all the Landau levels with $n\geq 1$ the kinematic constrain implies that the magnetic field
cannot be arbitrarily  large without violating it. This determines a
maximum value for the magnetic field, $q_f B_{\rm max} = (M^2 - 4
m_f^2)/8$, and for $B\geq B_{\rm max}$ we have that all terms in Eq.~(\ref{fulldecaysum})
with $n\geq 1$ vanishes and only the LLL terms contributes to the decay 
width\footnote{We thank G. Endr\H odi for explicitly pointing out an error in an earlier
version of these calculations and recalling us that the LLL for the decay into fermions
is special and survives in the decay width at large $B$.  We are here neglecting possible
  backreactions of the charged fermion fields on the scalar field,
  which can induce magnetic field corrections (as well as thermal
  corrections) to the scalar field and change this condition. These
  effects are of course beyond the one-loop approximation considered
  in this work.}. 

\section{The weak and strong magnetic field limits for the scalar field decay width from the Yukawa coupling}
\label{limitingcases}

Having computed the general expression for the decay width in the
previous section, let us now focus on the approximation for the decay
width in the cases of a strong and a weak magnetic field. These
limiting cases are usually considered in the literature, so it is
useful to analyze them as well for comparison.  By weak and strong
magnetic field here we mean $q_f B \ll m_f^2$ and $q_f B \gtrsim
m_f^2$, respectively.

\subsection{The weak magnetic field approximation}

To consider the weak magnetic field, let us first consider the Dirac
propagator in this case.  Expanding the exponential and tangent
functions in the expression Eq.~(\ref{schwinger_propertime}), we
immediately get $S_m(K)$ as a  series in powers of $q_fB$. Up to order
$(q_fB)^2$, it is expressed as
\begin{eqnarray}
S_{mw}(K)=\frac{\slashed{K}+m_f}{K^2-m_f^2}+q_fB\frac{i(\slashed{K}_\shortparallel+m_f)\gamma^1\gamma^2}{(K^2-m_f^2)^2}
+(q_fB)^2\left[\frac{2\slashed{K}_\perp}{(K^2-m_f^2)^3}-\frac{2K_\perp^2(\slashed{K}+m_f)}{(K^2-m_f^2)^4}\right].
\label{wfprop}
\end{eqnarray}
We can also write the above expression as
\begin{eqnarray}
S_{mw}(K)=F(K,m_f,m_i)\frac{1}{K^2-m_i^2}\Bigg|_{m_i=m_f},
\label{wfp_spec}
\end{eqnarray}
where 
\begin{eqnarray}
F=(\slashed{K}+m_f)+i \, a_j \,\left(\slashed{K}_\shortparallel +
m_f\right)\gamma^1\gamma^2 + b_j\slashed{K}_\perp +c_j K_\perp^2
(\slashed{K}+m_f),
\end{eqnarray}
with coefficients $a_j, b_j$ and $c_j$ carrying the derivative
operators,
\begin{eqnarray}
a_j=q_fB\frac{\partial}{\partial
  m_j^2},~~b_j=(q_fB)^2\frac{\partial^2}{\partial (m_j^2)^2},
~~c_j=-\frac{1}{3}(q_fB)^2\frac{\partial^3}{\partial (m_j^2)^3}.
\end{eqnarray}

The one-loop correlation function in this case is similarly given by
\begin{eqnarray}
i\Pi_w(P) &=& (ig)^2 \int\frac{d^4K}{(2\pi)^4}
\mathrm{Tr}\left[S_{mw}(-Q)S_{mw}(K)\right]\nonumber\\ &=& g^2 \int
\frac{d^4K}{(2\pi)^4} \left[{\cal N}_{a,b,c}
  \frac{1}{(Q^2-m_1^2)(K^2-m_2^2)}\right]\Bigg\vert_{m_1=m_2=m_f},
\label{Piweak}
\end{eqnarray}
where 
\begin{eqnarray}
{\cal N}_{a,b,c}= -\textrm{Tr}\left[F(K,m_f,m_2)~F(-Q,m_f,m_1)\right].
\label{TrFF}
\end{eqnarray} 
Here, the masses $m_1$ and $m_2$ are variables on which the mass
derivatives inside ${\cal N}_{a_j,b_j,c_j}$ act on in the  square
bracket term in Eq.~(\ref{Piweak}).  Now, proceeding similarly as in
the previous section, we can write down the imaginary part of the
one-loop scalar self-energy for the decay  process,
\begin{eqnarray}
\textrm{Im}~\Pi_w(P)  &=& \pi~g^2~\int \frac{d^3k}{(2\pi)^3}
\frac{{\cal
    N}_{a,b,c}(\omega_1,\omega_2)~(1-n_F(\omega_1)-n_F(\omega_2))}{4~\omega_1\omega_2}~\delta(p_0-\omega_1-\omega_2)
\end{eqnarray}
where we have defined
\begin{eqnarray}
\omega_1^2 &=& q^2 + m_1^2,\nonumber\\ \omega_2^2 &=& k^2 + m_2^2.\nonumber
\end{eqnarray}
Let us evaluate the expression of ${\cal N}_{a,b,c}$ up to
$\mathcal{O}(q_fB)^2$. {}First, consider the term inside the trace in
Eq.~(\ref{TrFF}),
\begin{eqnarray}
F(K,m_f,m_2)~F(-Q,m_f,m_1) &=&
-\left[(\slashed{Q}-m_f)+a_1~i~\left(\slashed{Q}_\shortparallel -
  m_f\right)\gamma^1\gamma^2 + b_1\slashed{Q}_\perp +c_1Q_\perp^2
  (\slashed{Q}-m_f)\right] \nonumber\\ &\times&
\left[(\slashed{K}+m_f)+a_2~i~\left(\slashed{K}_\shortparallel +
  m_f\right)\gamma^1\gamma^2 + b_2\slashed{K}_\perp +c_2K_\perp^2
  (\slashed{K}+m_f)\right].
\end{eqnarray}

So,
\begin{eqnarray}
{\cal N}_{a,b,c} &=&
-\textrm{Tr}\left[F(K,m_f,m_2)~F(-Q,m_f,m_1)\right] \nonumber\\ &=&
4\left(1+c_1Q_\perp^2+c_2K_\perp^2\right)\left(K\cdot Q -
m_f^2\right)+4a_1a_2\left((K\cdot
Q)_\shortparallel-m_f^2\right)\nonumber\\ && + ~4(b_1+b_2)\left(K\cdot
Q\right)_\perp.
\end{eqnarray}

Now neglecting the external transverse momentum $P_\perp$, in turn we
obtain 
\begin{eqnarray}
{\cal N}_{a,b,c}(\omega_1,\omega_2) &=& 4(1+a_1a_2+(c_1+c_2)K_\perp^2)
(K\cdot Q-m_f^2)-4(a_1a_2+b_1+b_2)K_\perp^2\nonumber\\ &=&
4\left(1+a_1a_2+\frac{2}{3}(c_1+c_2)k^2\right)(K\cdot
Q-m_f^2)-\frac{8}{3}(a_1a_2+b_1+b_2)k^2,
\end{eqnarray}
where we have also taken $K_\perp^2=k^2\left\langle \sin^2\theta\right\rangle_\theta=\frac{2}{3}k^2$, an approximation exploiting the choice of the frame which is valid in the weak field limit.  

So, the decay width in the rest frame of the decaying boson for the
weakly magnetized medium is given by
\begin{eqnarray}
\Gamma^{\textrm{w}} &=& \frac{\mathrm{Im}~\Pi_w}{M}\nonumber\\ &=&
\left[\frac{4\pi g^2}{M}(1+a_1a_2)~J_1+\frac{8\pi g^2}{3M}
  (c_1+c_2)~J_2-\frac{8\pi g^2}{3M}
  (a_1a_2+b_1+b_2)~J_3\right]\Bigg|_{m_1=m_2=m_f},
\label{fd_wfa_final}
\end{eqnarray}
where 
\begin{eqnarray}
J_1 &=&  \int\frac{d^3k}{(2\pi)^3}\left(K\cdot
Q-m_f^2\right)\frac{\left[1-n_F(\omega_1)-n_F(\omega_2)\right]}{4\omega_1\omega_2}
\delta(p_0-\omega_1-\omega_2),
\label{J1}
\\ J_2 &=&  \int\frac{d^3k}{(2\pi)^3}~k^2\left(K\cdot
Q-m_f^2\right)\frac{\left[1-n_F(\omega_1)-n_F(\omega_2)\right]}{4\omega_1\omega_2}
\delta(p_0-\omega_1-\omega_2),
\label{J2}
\\ J_3 &=&
\int\frac{d^3k}{(2\pi)^3}~k^2\frac{\left[1-n_F(\omega_1)-n_F(\omega_2)\right]}{4\omega_1\omega_2}
\delta(p_0-\omega_1-\omega_2).
\label{J3}
\end{eqnarray}
In the Appendix~\ref{AppB} we derive the respective expressions for
Eqs.~(\ref{J1}), (\ref{J2}) and (\ref{J3}).  Proceeding with the final
momentum integrations in the scalar field rest frame, we obtain that
\begin{equation}
\Gamma^{\textrm{w}}_M(T,B) =
\Gamma^{\textrm{w}}_M(T=0,B)\left[1-2n_F\left(\frac{M}{2}\right)\right],
\label{smallfield}
\end{equation}
where
\begin{eqnarray} 
\Gamma^{\textrm{w}}_M(T=0,B) &=& \frac{g^2
  M}{8\pi}\left(1-\frac{4m_f^2}{M^2}\right)^{3/2}
\Theta\left(M^2-4m_f^2\right) \nonumber \\ &\times& \left[ 1-
  \frac{2(q_fB)^2}{M^4} \left(\frac{8}{3} - \frac{2m_f^2}{M^2} \right)
  \left(1-\frac{4m_f^2}{M^2}\right)^{-2}\right].
\end{eqnarray}
Note that the first term on the right hand side of the above
expression can be identified as the usual vacuum decay width
$\Gamma^{\textrm{v}}_M$, when evaluated at $B=0$.

\subsection{The strong magnetic field approximation}

Let us now obtain the limiting case of the decay width in a strong
magnetic field, in particular when $q_f B \gg m_f^2$.  In presence of
a very strong magnetic field all the Landau levels with $n\ge 1$ are
pushed to infinity compared to the lowest Landau Level (LLL) with
$n=0$. So, for the case of strong external magnetic field we can
assume the  LLL approximation, by the virtue of which the fermion
propagator in Eq.~(\ref{decomposed_propagator}) reduces to a
simplified form as
\begin{eqnarray}
iS_{ms}(K)=ie^{-{K_\perp^2}/{q_fB}}~~\frac{\slashed{K}_\shortparallel+m_f}{
  K_\shortparallel^2-m_f^2}(1-i\gamma_1\gamma_2),
\label{prop_sfa}
\end{eqnarray}
where $K$ is four-momentum and we have used the properties of the
generalized Laguerre  polynomial, $L_n\equiv L_n^0$ and $L_{-1}^\alpha
= 0$. The one-loop scalar self-energy can then be written as
\begin{eqnarray}
i\Pi_{LLL}(P)&=& g^2 \int\frac{d^2K_\perp}{(2\pi)^2}
\exp\left(\frac{-K_\perp^2-Q_\perp^2}{q_fB}\right)
\nonumber\\ &\times&\int\frac{d^2K_\shortparallel}{(2\pi)^2}~\mathrm{Tr}
\left[\frac{\slashed{Q}_\shortparallel-m_f}{Q_\shortparallel^2-m_f^2}(1-i\gamma_1\gamma_2)
  \frac{\slashed{K}_\shortparallel+m_f}{K_\shortparallel^2-m_f^2}(1-i\gamma_1\gamma_2)\right].
\label{PiLLL}
\end{eqnarray}
Evaluation of the trace and the Gaussian integral over $K_\perp$ in
Eq.~(\ref{PiLLL}) is rather straightforward, which yields
\begin{eqnarray}
i\Pi_{LLL}(P) &=& g^2 \frac{q_fB}{\pi}
\exp\left(\frac{-P_\perp^2}{2q_fB}\right)\int\frac{d^2K_\shortparallel}{(2\pi)^2}~\frac{(K\cdot
  Q)_\shortparallel-m_f^2}{(Q_\shortparallel^2-m_f^2)(K_\shortparallel^2-m_f^2)},\nonumber\\ &=&
i g^2 \frac{q_fB}{\pi}
\exp\left(\frac{-P_\perp^2}{2q_fB}\right)~I_{1,LLL},
\end{eqnarray}
where $I_{1,LLL}$ is the same momentum integral we have already
computed in the Appendix~\ref{appA} and given by Eq.~(\ref{I1}) when
it is evaluated by considering only the LLL term, i.e., by taking
$m_f' \to m_f$ in there.   Hence, the imaginary part of the one-loop
scalar field self-energy in the LLL approximation becomes
\begin{eqnarray}
\mathrm{Im}~ \Pi_{LLL}(P) &=& g^2 \frac{q_fB}{\pi}
\exp\left(\frac{-P_\perp^2}{2q_fB}\right)~\mathrm{Im}~I_1(m_f' \to
m_f) \nonumber \\ &=& g^2\frac{q_fB}{2\pi}~
e^{\frac{-P_\perp^2}{2q_fB}}\left(1-\frac{4m_f^2}{P_\shortparallel^2}
\right)^{\frac{1}{2}}\Theta
\left(P_\shortparallel^2-4m_f^2\right)\Bigl[1-n_F(p_+^s)-n_F(p_-^s)\Bigr],
\end{eqnarray}
where 
\begin{eqnarray}
p_\pm^{s} = \frac{p_0}{2}\pm
\frac{p_3}{2}\sqrt{1-\frac{4m_f^2}{P_\shortparallel^2}}
\label{p_pm_s}
\end{eqnarray}
and the decay width in the rest frame of the decaying scalar, in
presence of a strong background magnetic field and in the LLL
approximation is given by 
\begin{eqnarray}
\Gamma_{M,LLL} &=& \frac{\mathrm{Im}~\Pi_{LLL}(P=M)}{M} \nonumber
\\ &=& g^2\frac{q_fB}{2\pi M}  ~ \left(1-\frac{4m_f^2}{M^2}
\right)^{{1}/{2}}\Bigl[1-2n_F\left(\frac{M}{2}\right)\Bigr]\Theta
\left(M^2-4m_f^2\right) .
\label{largefield}
\end{eqnarray}
Equation~(\ref{largefield}) can also be verified by putting $n=0$ in
Eq.~(\ref{fulldecay}).  

\section{Discussion of the results}
\label{results}

\begin{center}
\begin{figure}[!htb]
\subfigure[The weak field
  case]{\includegraphics[width=7cm]{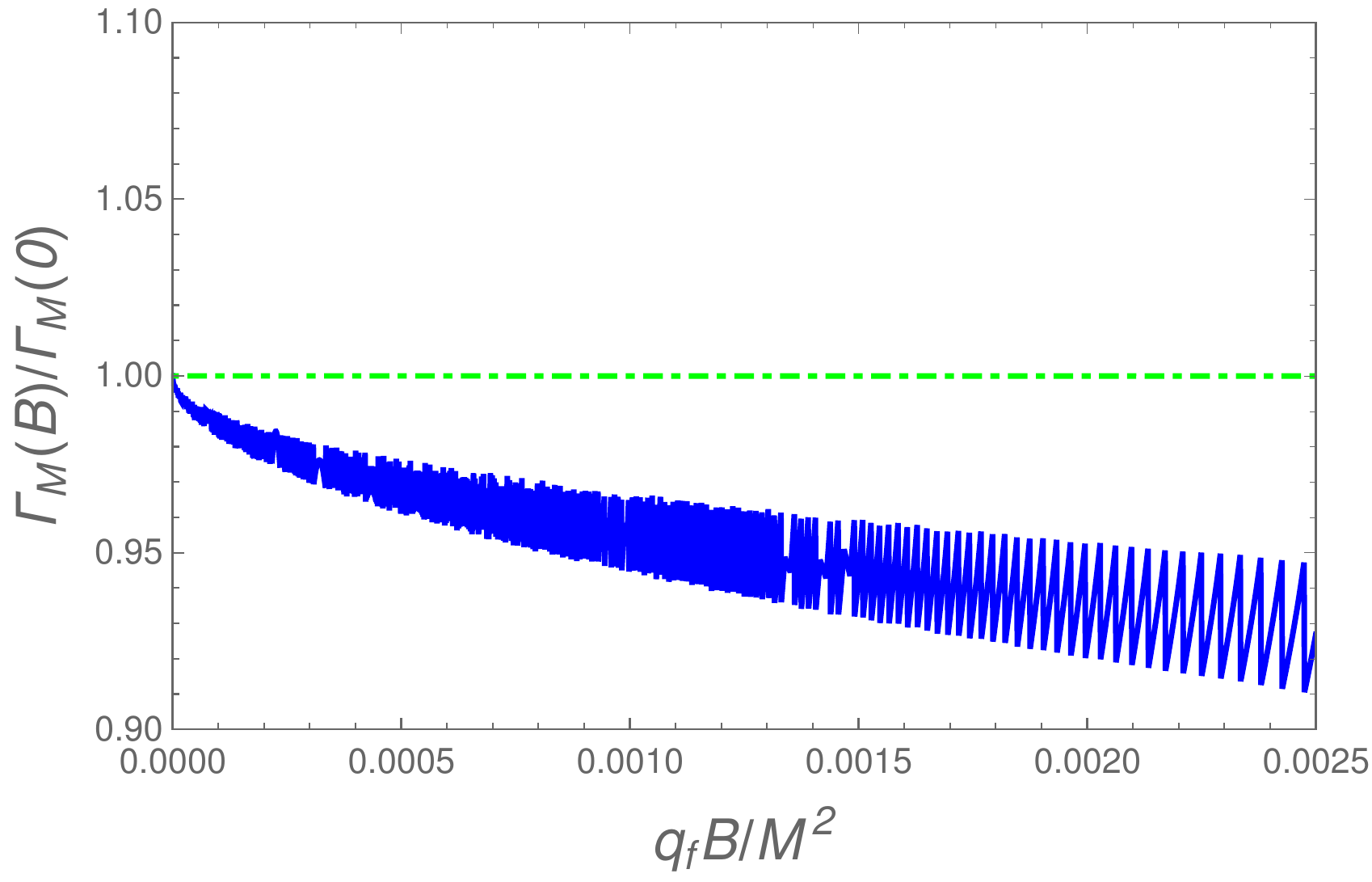}} \subfigure[The
  strong field case]{\includegraphics[width=6.7cm]{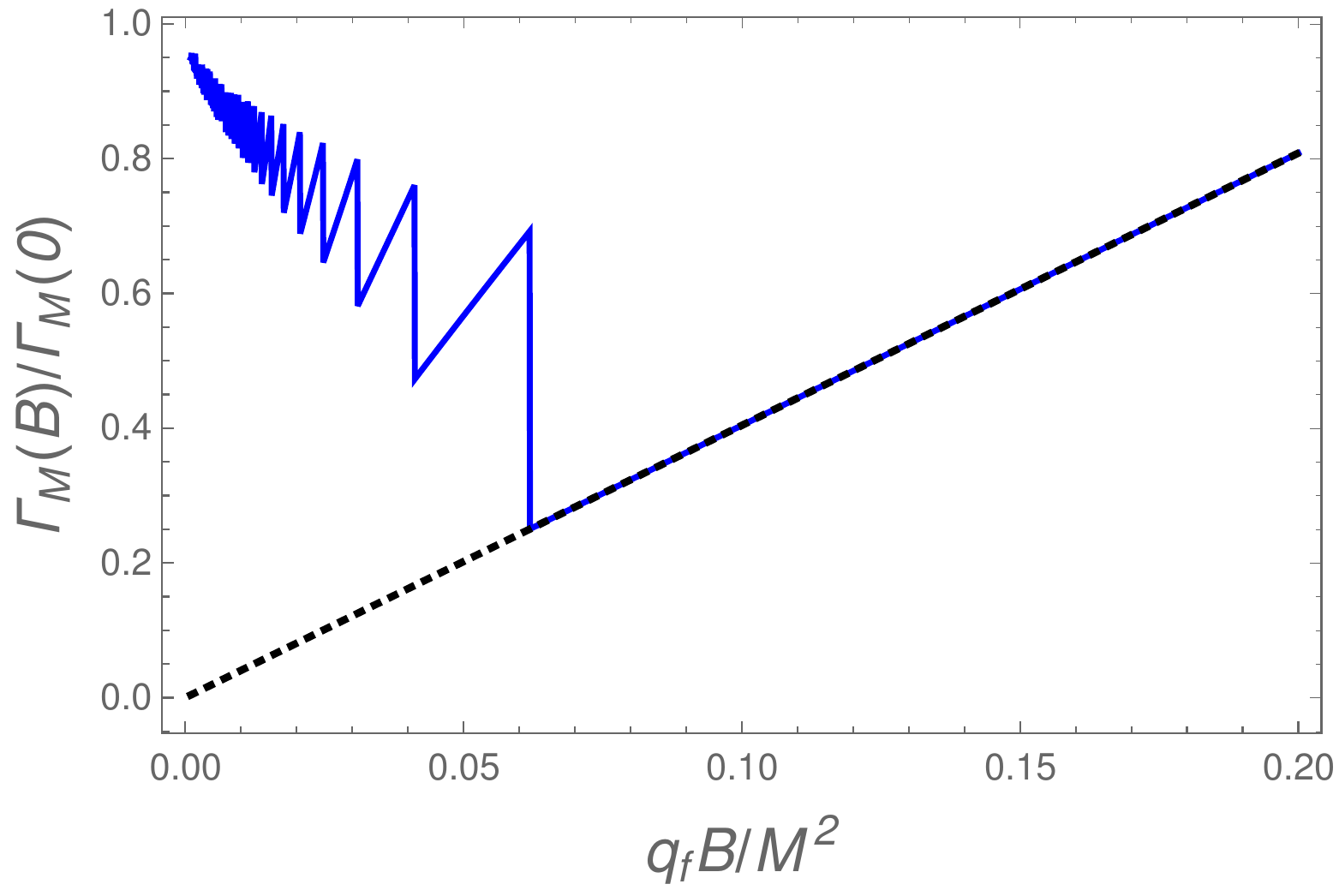}}
\caption{The decay width (normalized by the magnetic field independent
  result), as a function of the magnetic field. The solid lines in
  both (a) and (b) are the result of Eq.~(\ref{fulldecaysum}), the
  dotted line in (b) is the strong field result
  Eq.~(\ref{largefield})  and the dash-dotted line in (a) is the
  weak field approximation  Eq.~(\ref{smallfield}). We have set the
  fermion mass as $m_f=M/20$. }
\label{fig2}
\end{figure}
\end{center}

Let us now compare the results we have obtained for the decay width in
the previous sections.  We start by comparing the arbitrary field
result given by Eq.~(\ref{fulldecaysum}) with the two approximating
expressions from the latter section, i.e., the weak field
approximation  given by Eq.~(\ref{smallfield}) and the strong field
approximation given by Eq.~(\ref{largefield}).  This is shown in
{}Fig.~\ref{fig2}. {}For convenience, we normalize the results by the
decay width in the absence of an external magnetic field,

\begin{equation}
\Gamma_M(B=0) = \frac{g^2
  M}{8\pi}\left(1-\frac{4m_f^2}{M^2}\right)^{3/2}
\left[1-2n_F\left(\frac{M}{2}\right)\right]
\Theta\left(M^2-4m_f^2\right).
\label{zerofield}
\end{equation}

{}From the results of {}Fig.~\ref{fig2} we can see that the general
behavior of the decay width is to {\it decrease} with the magnetic
field.  The sharp teeth-saw behavior is consequence of the discretized
Landau levels in Eq.~(\ref{fulldecaysum}).  Within each Landau level,
the decay width tends to increase, till the kinematical constrain is
reached and we begin again with the next Landau level, which gives
origin to the teeth-saw behavior seen in {}Fig.~\ref{fig2}.  Note that
the highest Landau levels are populated initially at lowest values of
$B$, with the LLL populated for very last at the highest value of $B$,
before the decay width eventually vanishes for all Landau levels
with $n \geq 1$ due to the kinematic 
constrain, remaining only the LLL contribution.  The weak magnetic field also shows a decrease with the
magnetic field, though barely apparent in the scale of
{}Fig.~\ref{fig2}a, where we show it only up to its range of
validity. In particular, we see that at the value $q_f B = m_f^2$ the
weak field approximation already over estimates the decay width by
around $5\%$ compared to the arbitrary field result.  The strong field
approximation, expressed in Eq.~(\ref{largefield}) and shown as the dotted line in {}Fig.~\ref{fig2}b, has always an {\it increasing}
(linear in $B$) behavior with the magnetic field. 
This can
be better seen by noticing that the sum over the Landau levels in
Eq.~(\ref{fulldecaysum}) has an explicit analytic continuation in
terms of zeta-functions,
\begin{eqnarray}
\Gamma_M(B,T) &=& \frac{\sqrt{2}g^2\left(q_fB\right)^{3/2}}{\pi M^2}
 \sqrt{\frac{M^2-4m_f^2}{8 q_f B}} \Theta\left(M^2-4m_f^2\right)
\nonumber\\
&+&
\frac{\sqrt{2}g^2\left(q_fB\right)^{3/2}}{\pi M^2}
\left[ - \frac{3}{2}
  \zeta\left(-\frac{1}{2}, \frac{M^2-4m_f^2}{8 q_f B} \right) \right.
  \nonumber\\ &-& \left.  \frac{N_{\rm max}(B)}{2}
  \zeta\left(\frac{1}{2}, \frac{M^2-4m_f^2}{8 q_f B} \right) +
  \frac{3}{2} \zeta\left(-\frac{1}{2}\right) + \frac{M^2-4m_f^2}{16
    q_f B} \zeta\left(\frac{1}{2}\right)   \right] \nonumber
\\ &\times& \left[1-2n_F\left(\frac{M}{2}\right)\right]
\Theta\left[N_{\rm max}(B)-1\right],
\label{fulldecayzeta}
\end{eqnarray}   
where in the above expression we have again separated explicitly the LLL $n=0$
contribution, while explicitly summing over the $n\geq 1$ Landau levels and
\begin{equation}
\zeta(s,a) = \sum_{k=0}^{\infty} \frac{1}{(k+a)^s},
\label{zetafunc}
\end{equation}
is the Hurwitz zeta function~\cite{zeta}.

\begin{center}
\begin{figure}[!htb]
\includegraphics[width=7cm]{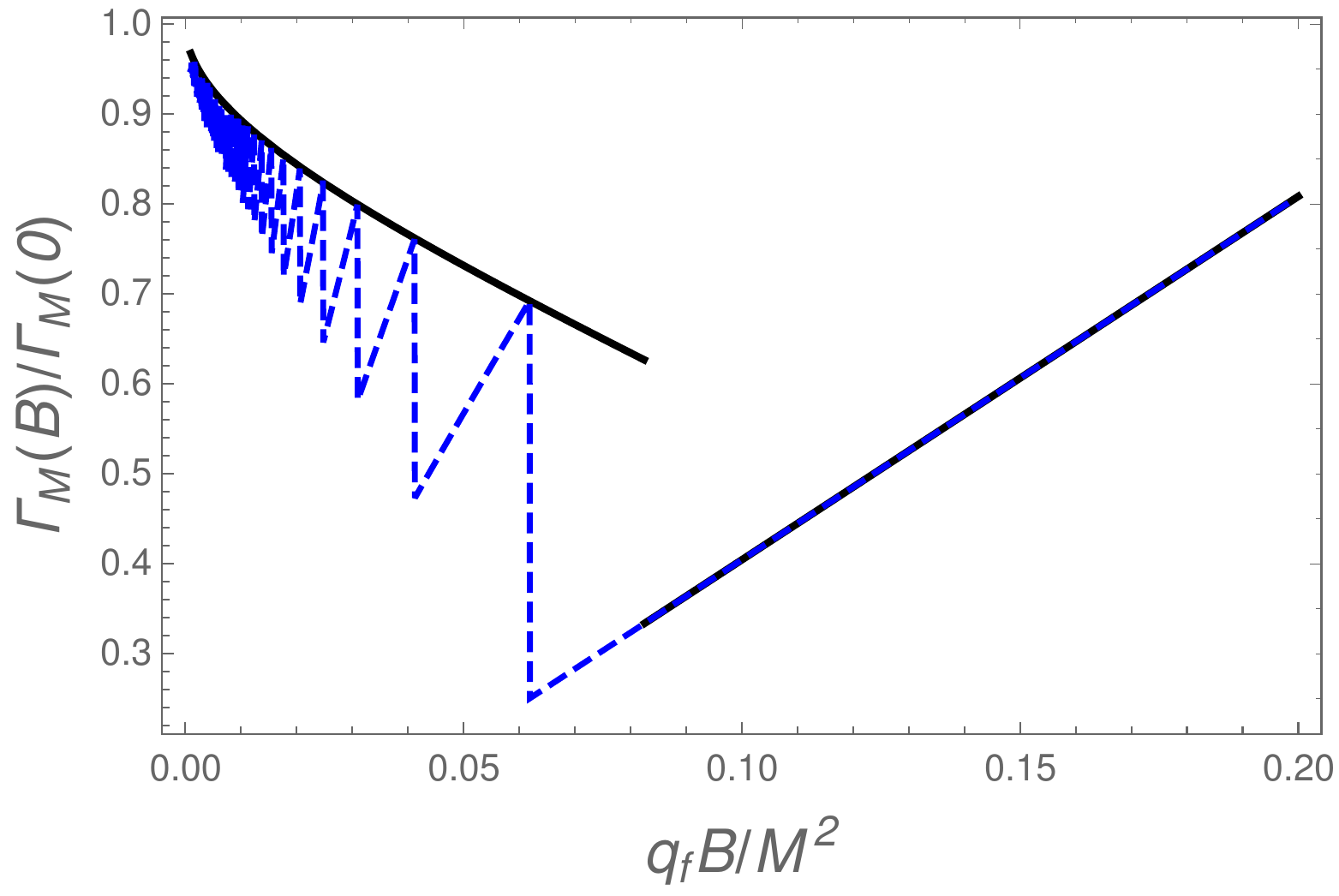}
\caption{The decay width (normalized by the magnetic field independent
  result), as a function of the magnetic field. The dashed line is the
  zeta-function analytic continuation given by
  Eq.~(\ref{fulldecayzeta}),  while the solid line is
  Eq.~(\ref{fulldecaysum}). We have set the fermion mass as
  $m_f=M/20$. }
\label{fig3}
\end{figure}
\end{center}

In {}Fig.~\ref{fig3} we compare the result given by
Eq.~(\ref{fulldecaysum}) with the analytic continuation of it in terms
of the zeta-functions.

The analytic continuation result given by Eq.~(\ref{fulldecayzeta})
makes it clearer that the decay width is always a decreasing function
with the magnetic field, except,
of course, at the threshold point, where the decay width contribution with Landau
levels with $n\geq 1$ vanishes and only remains the LLL contribution, that now grows
linearly with $B$.

\section{Bosonic decay in the presence of an arbitrary magnetic field}
\label{boson}

Let us now consider the case of decay of a heavy neutral scalar field
into a pair of charged scalars, $\Phi \to \chi + \chi^*$.  The
interacting Lagrangian density is given by the trilinear coupling
between the fields,
\begin{eqnarray}
\mathcal{L} = g\Phi\chi^*\chi.
\end{eqnarray}
The scalar field $\Phi$ self-energy is now given by
\begin{eqnarray}
i\Pi^b (P) = (ig)^2 \int \frac{d^4K}{(2\pi)^4} D_B(P-K) D_B(K),
\end{eqnarray}
where $D_B(K)$ is the bosonic propagator in presence of an arbitrary
external magnetic field and it is given by~\cite{Ayala:2004dx}
\begin{eqnarray}
D_B(K) = 2\sum\limits_{n=0}^\infty
\frac{(-1)^nL_n\left(\frac{2K_\perp^2}{q_b
    B}\right)~e^{-\frac{K_\perp^2}{q_b
      B}}}{K_\shortparallel^2-(2n+1)q_b B-m_b^2+i\epsilon},
\end{eqnarray}
where $q_b$ is the bosonic charge.
Incorporating this expression for the bosonic propagator and
neglecting the external transverse momentum  $P_\perp$ again, it
allows us to obtain an analytic result for the self-energy, which
becomes 
\begin{eqnarray}
i\Pi^b(P_\shortparallel) =
-4g^2\!\!\int\!\!\frac{d^2K_\shortparallel}{(2\pi)^2}\!\!\int\!\!\frac{d^2K_\perp}{(2\pi)^2}\!\!\sum_{l,n=0}^{+\infty}
\!\!\!(-1)^{l+n} e^{-\frac{2K_\perp^2}{q_b
    B}}L_l\!\left(\!\frac{2K_\perp^2}{q_b
  B}\!\right)\!L_n\!\left(\!\frac{2K_\perp^2}{q_b B}\!\right)\!
\frac{1}{(K_\shortparallel^2-m_b'^2)(Q_\shortparallel^2-m_b''^2)},
\end{eqnarray}
where we have used $Q \equiv P-K$ with $m_b'^2 = m_b^2+(2n+1)q_b B$
and $m_b''^2 = m_b^2+(2l+1)q_b B$. 

Now, using the same orthogonality relation as used in
Eq.~(\ref{laug_ortho}), we obtain
\begin{eqnarray}
\Pi^b(P) &=& 4ig^2\frac{q_b B}{8\pi} \sum_{n=0}^{+\infty}
\int\frac{d^2K_\shortparallel}{(2\pi)^2}
\frac{1}{(K_\shortparallel^2-m_b'^2)(Q_\shortparallel^2-m_b'^2)},
\label{Pib}
\end{eqnarray}
where now
\begin{equation}
\int\frac{d^2K_\shortparallel}{(2\pi)^2} \to i T \sum_j
\int\frac{dk_3}{2\pi},
\end{equation}
and the Minkowski time component of the momentum is replaced by  $k_0
\to i \omega_j$, where $\omega_j = 2 j \pi T$, $j=0,\pm 1,\ldots$, are
the Matsubara's frequencies for bosons.

Similarly as the fermionic loop derivation done in Appendix
\ref{appA}, we can perform the resulting Matsubara sum in the bosonic
loop by using  again the mixed  representation technique prescribed by
Pisarski~\cite{Pisarski:1987wc,Bandyopadhyay:2016fyd}, but this time
for bosons,
\begin{eqnarray}
\frac{1}{K_\shortparallel^2-m_b^2} \equiv \frac{1}{k_0^2-E_{k}^2} =
\int\limits_0^{1/T} d\tau e^{k_0\tau} \Delta_M(\tau,k),
\label{mixed_representationb}
\end{eqnarray}
where
\begin{eqnarray}
\Delta_M(\tau,k) =
\frac{1}{2E_{k}}\left\{\left[1+n_B(E_{k})\right]e^{-E_{k}\tau}+n_B(E_{k})e^{E_{k}\tau}\right\},
\end{eqnarray}
where we have defined the dispersion relation as $E_k^2 = k_3^2 +
m_b'^2$ and $n_B(E_k)$ in the above equation is the Bose-Einstein
distribution function.  This way, we get for the Masubara's sum for
the bosonic loop the result 
\begin{eqnarray}
T\sum_{k_0} \frac{1}{(k_0^2-E_{k}^2)(q_0^2-E_{q}^2)} =
\sum\limits_{r,l=\pm 1}
\frac{rl}{4E_{k}E_{q}}~\frac{1+n_B(rE_{k})+n_B(lE_{q})}{(p_0+rE_{k}+lE_{q})},
\label{loopB}
\end{eqnarray}
We now proceed similarly as in the derivation for the discontinuity of
the fermionic loop that determines the decay width.  Using
Eq.~(\ref{disc_delta}) to evaluate the discontinuity and being
interested only in the contribution from decay (and not Landau
damping), we can choose $r=l=-1$ in Eq.~(\ref{loopB}). Then, by also
using Eq.~(\ref{deltaf_prop}) to perform the $k_3$ integration, we
finally obtain that
\begin{eqnarray}
\mathrm{Im}~\Pi^b(P_\shortparallel) &=& \frac{g^2q_b B}{4\pi}
\sum_{n=0}^{+\infty} \Theta(P_\shortparallel^2-4m_b'^2)
\left(1-\frac{4m_b'^2}{P_\shortparallel^2}\right)^{-\frac{1}{2}}
\frac{1}{P_\shortparallel^2} \left[1+n_B(p_+^b)+n_B(p_-^b)\right],
\end{eqnarray}
where 
\begin{eqnarray}
p_\pm^{b} = \frac{p_0}{2}\pm
\frac{p_3}{2}\sqrt{\left(1-\frac{4m_b'^2}{P_\shortparallel^2}\right)}.
\label{p_pm_b}
\end{eqnarray}

Thus, the one-loop decay width for the process $\Phi\to \chi+\chi^*$,
in the rest frame of the decaying heavy scalar, becomes
\begin{eqnarray}
\Gamma_M^b(B,T) &=& \frac{\mathrm{Im}~\Pi(P_\shortparallel=M)}{M}
\nonumber\\ &=& \frac{g^2q_b B}{4\pi M^3} \sum_{n=0}^{+\infty}
\Theta(M^2-4m_b'^2) \left(1-\frac{4m_b'^2}{M^2}\right)^{-\frac{1}{2}}
\left[1+2n_B\left(\frac{M}{2}\right) \right].
\label{decayb}
\end{eqnarray}
As in the fermionic loop case, the kinematic constrain in
Eq.~(\ref{decayb}) limits the upper value for the Landau levels such
that
\begin{equation}
n < {\rm Integer}\left[\frac{M^2 - 4 m_b^2}{8 q_b
    B}-\frac{1}{2}\right] \equiv N_{\rm max,b}(B),
\label{conditionb}
\end{equation}
and Eq.~(\ref{decayb}) becomes

\begin{eqnarray}
\Gamma_M^b(B,T) &=&\frac{g^2q_b B}{4\pi M^3} \sum_{n=0}^{N_{\rm
    max,b}(B)-1}
\left\{1-\frac{4\left[m_b^2+(2n+1)q_bB\right]}{M^2}\right\}^{-\frac{1}{2}}
\left[1+2n_B\left(\frac{M}{2}\right) \right].
\label{decaybsum}
\end{eqnarray}
The above expression also has an analytic continuation in terms of
zeta-functions, given by
\begin{eqnarray}
\Gamma_M^b(B,T) &=&\frac{\sqrt{2}g^2 \sqrt{q_b B}}{16\pi M^2}  \left[
  \zeta\left(\frac{1}{2} \right)-  \zeta \left( \frac{1}{2}, \frac{M^2
    - 4 m_b^2}{8 q_b B}+\frac{1}{2} \right)\right]
\left[1+2n_B\left(\frac{M}{2}\right) \right]\Theta\left(N_{\rm
  max,b}(B)-1\right).
\label{decaybzeta}
\end{eqnarray}
The above expression for the arbitrary magnetic field can be compared
with the corresponding limiting results for the bosonic decay
width. In the weak field approximations, which was derived in details
in Ref.~\cite{Piccinelli:2017yvl}, the decay width is given
by\footnote{Note that there is a sign misprint in Eq.~(21) of
  Ref.~\cite{Piccinelli:2017yvl}.}
\begin{eqnarray}
\Gamma_{M,{\rm weak}}^b(B,T) &\simeq&\frac{g^2}{16\pi M^2}
\sqrt{1-4\frac{m_b^2}{M^2}} \left[1+2n_B\left(\frac{M}{2}\right)
  \right] \Theta(M^2 - 4 m_b^2) \nonumber \\ &\times&  \left[1-
  \frac{2 (q_b B)^2}{3M^4} \left( 1-40
  \frac{m_b^2}{M^2}\right)\right].
\label{decaybweak}
\end{eqnarray}
Likewise, we can easily determine the strong field approximation,
given by the LLL contribution,
\begin{eqnarray}
\Gamma_{M,LLL}^b(B,T) &=&\frac{g^2q_b B}{4\pi M^3}
\left[1-\frac{4\left(m_b^2+q_b B\right)}{M^2}\right]^{-\frac{1}{2}}
\left[1+2n_B\left(\frac{M}{2}\right) \right]\Theta(M^2 - 4 m_b^2),
\label{decaybLLL}
\end{eqnarray}
and also subject to the range of applicability (following from
Eq.~(\ref{conditionb}) and that $m_b^2 < q_bB$),
\begin{equation}
1-4\frac{m_b^2+q_bB}{M^2} > 8 \frac{q_bB}{M^2} > 8 \frac{m_b^2}{M^2}.
\label{conditionLLL2}
\end{equation}
The largest range still comes when we set the boson mass to zero, with
the magnetic field then limited by the maximum value $q_b B < M^2/12$,
thus, smaller than the equivalent condition found for  the case of the
decay into fermions.

\begin{center}
\begin{figure}[!htb]
\subfigure[]{\includegraphics[width=7cm]{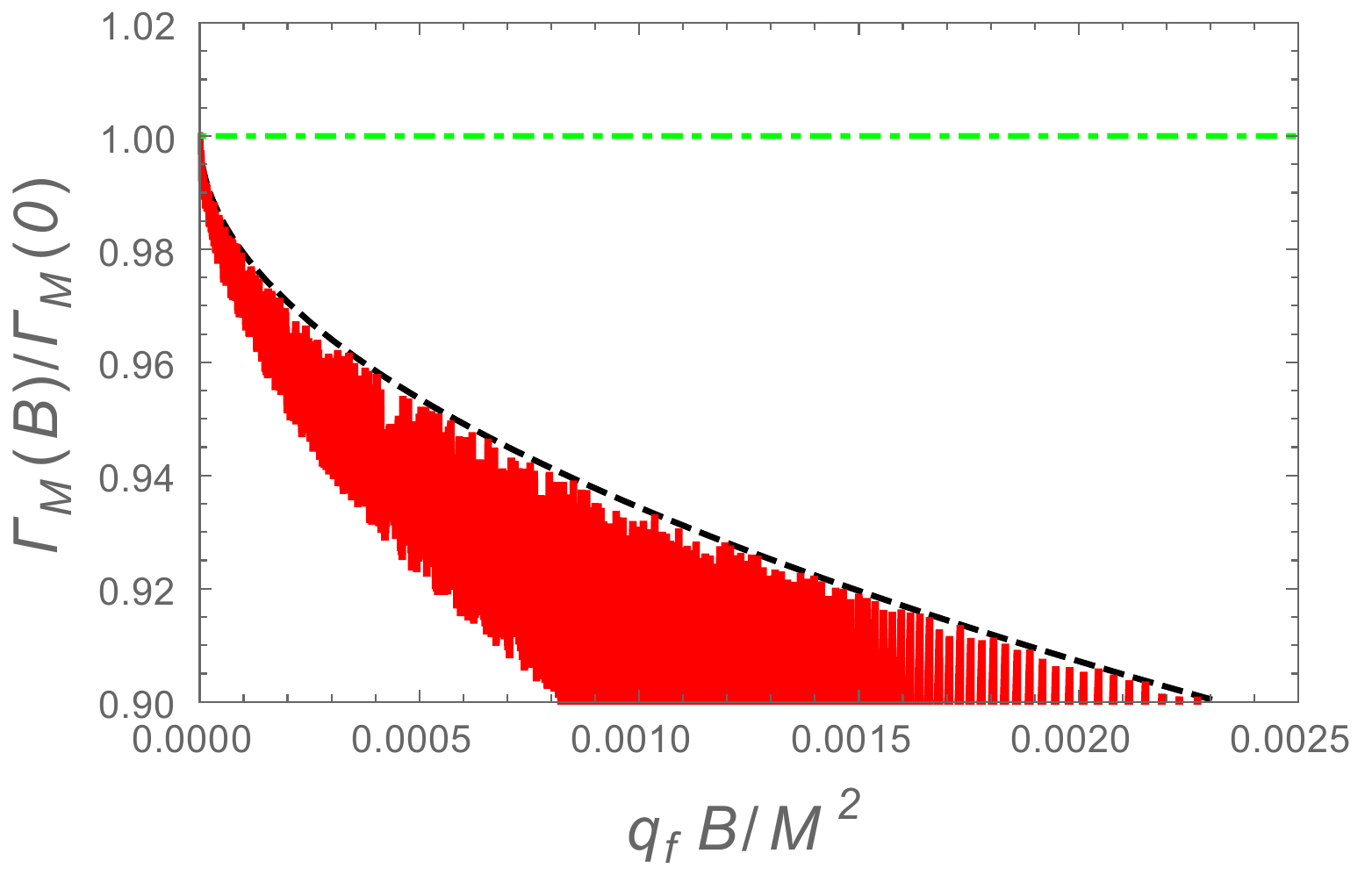}}
\subfigure[]{\includegraphics[width=6.7cm]{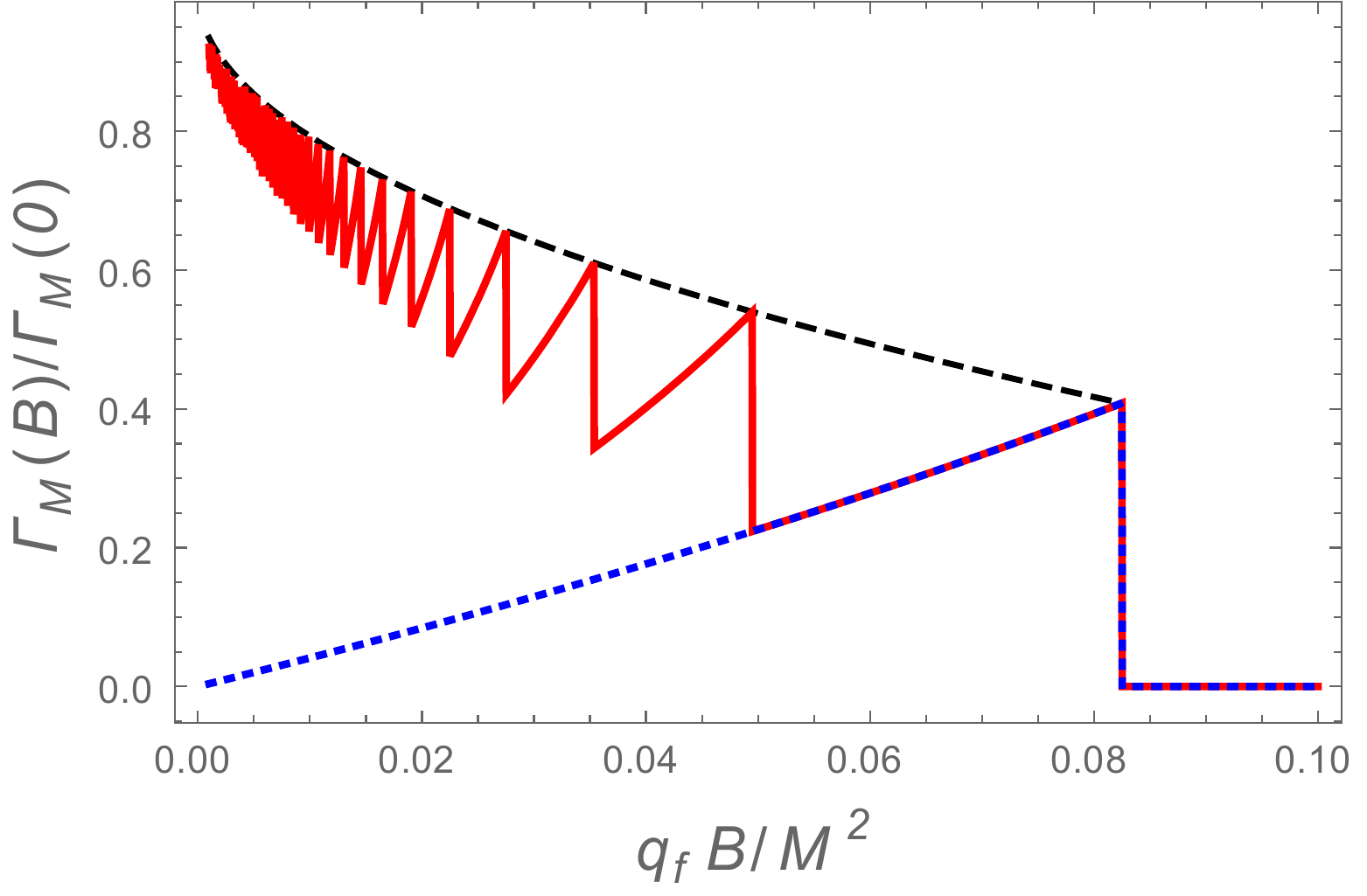}}
\caption{The decay width (normalized by the magnetic field independent
  result), as a function of the magnetic field for the case of a decay
  of a heavy neutral scalar field into charged ones.  The dashed line
  is the result of Eq.~(\ref{decaybzeta}), the solid line is given by
  Eq.~(\ref{decaybsum}),     the dash-dotted line is the weak field
  result given Eq.~(\ref{decaybweak}) and the dotted line is the
  strong field approximation  Eq.~(\ref{decaybLLL}). We have set the
  charged boson mass as $m_b=M/20$. }
\label{fig4}
\end{figure}
\end{center}

In {}Fig.~\ref{fig4} we compare the different results shown above. As
in the fermionic decay case, we have normalized the decay width by the
zero magnetic field result, given by
\begin{eqnarray}
\Gamma_M^b(B=0,T) &=& \frac{g^2}{16\pi M}
\left(1-\frac{4m_b^2}{M^2}\right)^{1/2}\left[1+2n_B\left(\frac{M}{2}\right)\right]\Theta
\left(M^2-4m_b^2\right).
\end{eqnarray}

In {}Fig.~\ref{fig4}, the sharp teeth-saw behavior is again
consequence of the discrete Landau levels considered in
Eq.~(\ref{decaybsum}).  We also note in {}Fig.~\ref{fig4}a that the
weak field approximation now over estimates the exact result by a much
larger percentage, as compared, e.g., to the decay into fermions
case. {}Furthermore, we see from  {}Fig.~\ref{fig4}b that, as in the
fermionic decay case, the LLL approximation predicts always an
increasing decay width with the magnetic field, while the analytic
full result is always decreasing with the magnetic field.
Contrary to the case of the decay into fermions where at large magnetic
fields only the LLL contributions survives, in the case of decay into
bosonic particles the kinematic constrain forces even the LLL term
to vanish. This is simply a consequence that the dispersion 
relation for bosons depends explicitly on $B$ even when $n=0$.

\section{Conclusions}
\label{conclusions}

In this work, we have studied the decay channels of a heavy neutral
scalar field into a pair of fermion-antifermion and a pair of charged
scalars when in the presence of an external magnetic field. Our
results indicate that there are similarities in these two decay
channels as a function of the magnetic field. In both the cases, we
observe that the decay width always tends to decrease with the
increase in the magnetic field up to the point where the LLL is filled. 
In a sense, this behavior could be
anticipated by the fact that in presence of a magnetic field the
effective mass of the decay products (as perceived by the decaying
particle) increases with the intensity of the magnetic field. This can
be seen easily when we look at the dispersion relation for a charged
scalar for example.

We have compared our analytical results, which can also be expressed
in an analytic continuation using zeta-functions, with the
approximated results in the limiting regimes, i.e., in the weak field
approximation (where the magnetic field satisfies $qB\ll m^2$) and in
the strong field or LLL approximation, where $qB \geq m^2$. 
We have seen that the kinematic condition leads to a
natural constrain in the upper magnitude of the external magnetic
field, beyond which the contribution of all the Landau levels with
$n\geq 1$ to decay width vanishes. In the case of decay into fermions
only the LLL contribution remains and make the decay width to grow linearly
with $B$. {}For the decay into bosons, however, even the LLL vanishes
as a consequence of the kinematic constrain and that it now depends explicitly
on $B$ even for the LLL. When contrasting both
results with the analytic expressions, we see that the weak field
approximation tends to over estimate the decay width. The strong field
approximation, on the other hand, leads to a complete wrong behavior
of the decay width in the decay into bosons case, showing a monotonic increase with $B$.  These
results point out that the use of these approximations in the
literature should be seen with quite some reservations.

We find that the decay width in strong magnetic fields is entirely blocked
in the case of decay into bosons.
This behavior
is different to what we see from the behavior of the decay widths when
in the presence of temperature or chemical potential. {}For instance,
the dependence on the temperature seen in the case of the decay into
fermions, given by Eq.~(\ref{fulldecaysum}), is a consequence of the
decrease of the phase space for the decay process as we increase the
temperature (i.e., Pauli blocking).  In the case of the decay into
bosons, the phase space increases due the bosonic nature of the
statistics, thus increasing the decay width with the temperature.  The
decrease of the decay width with the magnetic field and its blocking
at larger values of the external field is essentially a consequence of
the kinematic condition given its explicitly dependence on $B$.  
It would be interesting to study the case
where higher-loop effects are accounted for, or when the decaying
particle is also charged under the external magnetic field. In this
case, the dressing of the masses by magnetic field dependent terms
might lead to nontrivial effects.

The results obtained in this work can find many different
applications.  {}For instance the blocking of the decay in strong
magnetic field can have many important consequences. Possible
applications can be found in the context of early Universe
 where conditions predict the presence of extreme magnetic fields~\cite{Vachaspati:1991nm,Giovannini:2003yn,Piccinelli:2014dya}.
In particular, the study of how external conditions might affect decay
widths are of particular importance to understand the dynamics that
might be in play in cosmology~\cite{BasteroGil:2010pb}.  It is also
known that the presence of an external magnetic field can influence
the order of phase transitions~\cite{Ayala:2004dx,Duarte:2011ph}.  By
also influencing the decay processes happening in these phase
transitions, this can be potentially important in the problem of
baryogenesis in the early Universe.  In addition to this, another
situation that our results can be applicable is the study of the
decay processes following a heavy-ion collision, or, same in the
presence of the extreme fields in magnetars.  In heavy-ion collision
experiments, our results can be applicable in the study of the decay
of the neutral pion into quarks.  Our findings can be applicable and
be also of relevance in the study of the  processes involving the
decay of the Higgs into charged leptons, or, for the case of extensions
of the Standard Model, in the study of the decay of other scalar
particles into charged ones.

\appendix

\section{The fermion loop term}
\label{appA}

We start this section by evaluating the trace term appearing in
Eq.~(\ref{arbit_self_initial_expression}).  It can be expressed in the
form
\begin{eqnarray}
\textsf{Tr}\left[D_n(q_fB,-Q)~ D_m(q_fB,K)\right] = T_1 + T_2 +T_3, 
\end{eqnarray}
where the terms $T_1$, $T_2$ and $T_3$ are defined as
\begin{eqnarray}
T_1 &=&
L_n\left(\frac{2Q_\perp^2}{q_fB}\right)L_m\left(\frac{2K_\perp^2}{q_fB}\right)
\nonumber\\ &\times&\textsf{Tr}\Bigl[-\left(q_0\gamma_0-q_3\gamma_3-m_f\right)
  \left(1-i\gamma_1\gamma_2\right)\left(k_0\gamma_0-k_3\gamma_3+m_f\right)\left(1-i\gamma_1\gamma_2\right)\Bigr]\nonumber\\ &=&
-8\left[(K\cdot Q)_\shortparallel -m_f^2\right]
L_n\left(\frac{2Q_\perp^2}{q_fB}\right)L_m\left(\frac{2K_\perp^2}{q_fB}\right),
\\ T_2 &=&
L_{n-1}\left(\frac{2Q_\perp^2}{q_fB}\right)L_{m-1}\left(\frac{2K_\perp^2}{q_fB}\right)
\nonumber\\ &\times&\textsf{Tr}\Bigl[-\left(q_0\gamma_0-q_3\gamma_3-m_f\right)
  \left(1+i\gamma_1\gamma_2\right)\left(k_0\gamma_0-k_3\gamma_3+m_f\right)\left(1+i\gamma_1\gamma_2\right)\Bigr]\nonumber\\ &=&
-8\left[(K\cdot Q)_\shortparallel -m_f^2\right]
L_{n-1}\left(\frac{2Q_\perp^2}{q_fB}\right)L_{m-1}\left(\frac{2K_\perp^2}{q_fB}\right),
\\ T_3 &=& 16
L_{n-1}^1\left(\frac{2Q_\perp^2}{q_fB}\right)L_{m-1}^1\left(\frac{2K_\perp^2}{q_fB}\right)
\textsf{Tr}\Bigl[-\left(q_1\gamma_1+q_2\gamma_2\right)\left(k_1\gamma_1+k_2\gamma_2\right)\Bigr]\nonumber\\ &=&
64(K\cdot Q)_\perp
L_{n-1}^1\left(\frac{2Q_\perp^2}{q_fB}\right)L_{m-1}^1\left(\frac{2K_\perp^2}{q_fB}\right).
\end{eqnarray}
Here we note that the cross terms appearing in the expression of $D_n(q_fB,-Q)~ D_m(q_fB,K)$, e.g. $L_n()~L_{m-1}()$ or $L_n()~L^1_{m-1}()$, vanish while evaluating the trace due to the properties of the associated gamma matrices. In the following, we also work with the case of a vanishing external
transverse momentum, $P_\perp=0$, such that the expression for the
self-energy Eq.~(\ref{arbit_self_initial_expression}) then becomes
\begin{eqnarray}
\Pi(P_\shortparallel) &=& -ig^2
\int\frac{d^2K_\shortparallel}{(2\pi)^2}\int\frac{d^2K_\perp}{(2\pi)^2}~
e^{-\frac{2K_\perp^2}{q_fB}}\sum_{n=0}^\infty\sum_{m=0}^\infty
\left(\frac{(-1)^{m+n}}{Q_\shortparallel^2-m_f^2-2nq_fB}\right)\left(\frac{1}{K_\shortparallel^2-m_f^2-2mq_fB}
\right)\nonumber\\ && \times  \Biggl[8\left((K\cdot
  Q)_\shortparallel-m_f^2\right)\left\{L_n\left(\frac{2K_\perp^2}{q_fB}\right)L_m\left(\frac{2K_\perp^2}{q_fB}\right)+L_{n-1}
  \left(\frac {
    2K_\perp^2}{q_fB}\right)L_{m-1}\left(\frac{2K_\perp^2}{q_fB}\right)\right\}\nonumber\\ &&+64K_\perp^2
  L_{n-1}^1\left(\frac{2K_\perp^2}{q_fB}\right)L_{m-1}^1\left(\frac{2K_\perp^2}{q_fB}\right)\Biggr].
\end{eqnarray}
Now, to integrate the transverse part out, we use the orthogonality
relation of the Laguerre polynomials,
\begin{eqnarray}
\int\limits_0^\infty x^\alpha e^{-x} L_n^\alpha(x) L_m^\alpha(x) dx =
\frac{\Gamma(n+\alpha+1)}{n!}\delta_{n,m}.
\end{eqnarray}
Thus, we have that
\begin{eqnarray}
\int\limits_0^\infty e^{-\frac{2K_\perp^2}{q_fB}}
L_n\left(\frac{2K_\perp^2}{q_fB}\right)L_m\left(\frac{2K_\perp^2}{q_fB}\right)\frac{d^2K_\perp}{(2\pi)^2}
&=&
\frac{q_fB}{8\pi}~\delta_{n,m},\label{laug_ortho}\\ \int\limits_0^\infty
K_\perp^2~e^{-\frac{2K_\perp^2}{q_fB}}
L_{n-1}^1\left(\frac{2K_\perp^2}{q_fB}\right)L_{m-1}^1\left(\frac{2K_\perp^2}{q_fB}\right)\frac{d^2K_\perp}{(2\pi)^2}
&=& \frac{(q_fB)^2}{16\pi}~n \,\delta_{n,m}.
\end{eqnarray}
After integrating out the transverse part using these orthogonality
relations and using the Kronecker delta-functions  we are then left
with the result for the self-energy,
\begin{eqnarray}
\Pi(P_\shortparallel) &=&  -ig^2
\int\frac{d^2K_\shortparallel}{(2\pi)^2}\left[
  \frac{q_fB}{\pi}\sum_{n=0}^{+\infty} (2-\delta_{n,0}) \frac{ (K\cdot
    Q)_\shortparallel-m_f'^2}{(Q_\shortparallel^2-m_f'^2)(K_\shortparallel^2-m_f'^2)}
  \right. \nonumber\\ && \left. ~~~~~~~~~~+ 8\frac{(q_fB)^2}{\pi}
  \sum_{n=1}^{+\infty}\frac{n}{(Q_\shortparallel^2-m_f'^2)(K_\shortparallel^2-m_f'^2)}
  \right],
\label{self_energy_arbit_final}
\end{eqnarray}
where we have defined $m_f'^2 = m_f^2+2nq_fB$.

Let us call the first momentum integral in
Eq.~(\ref{self_energy_arbit_final}) as
\begin{equation}
I_1(P_\shortparallel)= \int\frac{d^2K_\shortparallel}{(2\pi)^2} \frac{
  (K\cdot
  Q)_\shortparallel-m_f'^2}{(Q_\shortparallel^2-m_f'^2)(K_\shortparallel^2-m_f'^2)},
\label{I1}
\end{equation}
and the second momentum  integral in
Eq.~(\ref{self_energy_arbit_final}) as
\begin{equation}
I_2(P_\shortparallel)= \int\frac{d^2K_\shortparallel}{(2\pi)^2} \frac{
  1}{(Q_\shortparallel^2-m_f'^2)(K_\shortparallel^2-m_f'^2)}.
\label{I2}
\end{equation}
At finite temperature, we can replace the momentum integral in the
above equations by

\begin{equation}
\int\frac{d^2K_\shortparallel}{(2\pi)^2} \to i T \sum_j
\int\frac{dk_3}{2\pi},
\end{equation}
and the Minkowski time component of the momentum is replaced by  $k_0
\to i \omega_j$, where $\omega_j = (2 j+1) \pi T$, $j=0,\pm 1,\ldots$,
are the Matsubara's frequencies for fermions.

Working first with the momentum integral $I_1$, we can now perform the
Matsubara sum in $I_1$ by using the mixed  representation technique
prescribed by
Pisarski~\cite{Pisarski:1987wc,Bandyopadhyay:2016fyd}. In this
prescription, we have that
\begin{eqnarray}
\frac{1}{K_\shortparallel^2-m_f'^2} \equiv \frac{1}{k_0^2-E_{k}^2} =
\int\limits_0^{1/T} d\tau e^{k_0\tau} \Delta_M(\tau,k),
\label{mixed_representation}
\end{eqnarray}
where
\begin{eqnarray}
\Delta_M(\tau,k) =
\frac{1}{2E_{k}}\left\{\left[1-n_F(E_{k})\right]e^{-E_{k}\tau}-n_F(E_{k})e^{E_{k}\tau}\right\},
\end{eqnarray}
where we have defined the dispersion relation as $E_k^2 = k_3^2 +
m_f'^2$ and $n_F(E_k)$ in the above equation is the {}Fermi-Dirac
distribution function.  Using this technique, we obtain for $I_1$ the
result
\begin{eqnarray}
I_1= \int\frac{dk_3}{2\pi} \sum_{l,r=\pm 1}\!\!
\frac{[(rl)E_{k}E_{q}-k_3q_3-m_f'^2]\left[1-n_F(rE_{k})\right]\left[1-n_F(lE_{q})\right]}{4(rl)E_{k}E_{q}(p_0-rE_{k}-lE_{q})}
\left[e^{-\beta(rE_{k}+lE_{q})}-1\right] .
\end{eqnarray}

The contribution of $I_1$ to the decay width is through its imaginary
part, which is computed as follows.  The imaginary part of $I_1$ is
extracted by using the following identity to  evaluate the
discontinuity,
\begin{eqnarray}
\textsf{Disc~}\left[\frac{1}{\omega +\sum\limits_i E_i}\right]_\omega
= - \pi\delta(\omega + \sum_i E_i).
\label{disc_delta}
\end{eqnarray}
After some straightforward algebra, we obtain the result,
\begin{eqnarray}
\mathrm{Im}~I_1 = \pi\!\!\int\!\!\frac{dk_3}{2\pi}\!\!\sum_{l,r=\pm
  1}\!\!\frac {[(rl)E_{k}E_{q}-k_3q_3-m_f^2]
  \left[1-n_F(rE_{k})-n_F(lE_{q})\right]}
       {4(rl)E_{k}E_{q}}~\delta(p_0-rE_{k}-lE_{q}),
\label{ImI1}
\end{eqnarray}

The Dirac delta-function in Eq.~(\ref{ImI1}) with two different values
of $r$ and $l$ represents four different process~\cite{Bellac:2011kqa}
.  The process with $r=-1$ and $l=-1$ violates the energy conservation
and, hence, it is disallowed. The processes $r=1, l=-1$ and $r=-1,
l=1$ signifies energy exchanges between the external heavy boson and
either one of the fermion/anti-fermion. These processes, in turn,
represent Landau damping. {}Finally, the process with $r=l=1$ clearly
shows that the energy of the external boson is decayed into the
fermion-antifermion pair. As in the present work we are interested
mainly in the decay width only, we then work with the case $r=l=1$,
yielding 
\begin{eqnarray}
\mathrm{Im}~I_1 &=&
\pi\int\frac{dk_3}{2\pi}\frac{[E_{k}E_{q}-k_3q_3-m_f^2]\left[1-n_F(E_{k})-n_F(E_{q})\right]}
            {4E_{k}E_{q}}~\delta(p_0-E_{k}-E_{q}).
\label{decayI1}
\end{eqnarray} 
The integral over $k_3$ in the above expression can now be performed
using the following property of the Dirac delta-function,
\begin{eqnarray}
\int\limits_{-\infty}^{\infty} dp_3~ f(p_3)~ \delta[g(p_3)] = \sum_{r}
\frac{f(p_{zr})}{\vert g^\prime(p_{zr})\vert}, 
\label{deltaf_prop}
\end{eqnarray}
where the zeros of the argument inside the Dirac delta-function are
denoted by $p_{zr}$.  Using Eq.~(\ref{deltaf_prop}) we can now perform
the $k_3$ integral in Eq.~(\ref{decayI1}) to obtain
\begin{eqnarray}
\mathrm{Im}~I_1&=& \frac{1}{2}
\left(1-\frac{4m_f'^2}{P_\shortparallel^2}
\right)^{{1}/{2}}\Bigl[1-n_F(p_+)-n_F(p_-)\Bigr] \Theta
\left(P_\shortparallel^2-4m_f'^2\right),
\label{finalI1}
\end{eqnarray}
where 
\begin{eqnarray}
p_\pm = \frac{p_0}{2}\pm
\frac{p_3}{2}\sqrt{1-\frac{4m_f'^2}{P_\shortparallel^2}}.
\end{eqnarray}

{}Following a similar procedure used to derive Eq.~(\ref{finalI1}), we
obtain for the second momentum integral $I_2$, given by
Eq.~(\ref{I2}), the result
\begin{eqnarray}
\mathrm{Im}~I_2&=& \frac{1}{4 P_\shortparallel^2}
\left(1-\frac{4m_f'^2}{P_\shortparallel^2}
\right)^{-{1}/{2}}\Bigl[1-n_F(p_+)-n_F(p_-)\Bigr] \Theta
\left(P_\shortparallel^2-4m_f'^2\right).
\label{finalI2}
\end{eqnarray}

{}Finally, we can write down the imaginary part of the fermion-loop
contribution to the heavy scalar particle as
\begin{eqnarray}
\mathrm{Im}~\Pi(P_\shortparallel) &=&  g^2 \frac{q_f B}{2\pi}
\sum_{n=0}^{+\infty} \left\{ (2-\delta_{n,0})
\left(1-\frac{4m_f'^2}{P_\shortparallel^2} \right)^{{1}/{2}} + 4\,n\,
\frac{q_f B}{ P_\shortparallel^2}
\left(1-\frac{4m_f'^2}{P_\shortparallel^2} \right)^{-{1}/{2}} \right\}
\nonumber \\ &\times & \Bigl[1-n_F(p_+)-n_F(p_-)\Bigr] \Theta
\left(P_\shortparallel^2-4m_f'^2\right).
\end{eqnarray}

\section{Momentum integrals in the case of the weak magnetic field limit}
\label{AppB}

Let us derive here the expressions for the momentum integrals in
Eqs.~(\ref{J1}), (\ref{J2}) and (\ref{J3}).  {}First, note that
\begin{eqnarray}
\omega_2^2 &=& k^2 + m_2^2,\nonumber\\ \therefore dk &=&
\frac{\omega_2 d\omega_2}{k},\nonumber\\ \omega_1^2 &=& p^2 +k^2 -
2pkc + m_1^2, ~~ (c=\cos\theta)\nonumber\\ \therefore dc &=& -
\frac{\omega_1 d\omega_1}{pk}.\nonumber
\end{eqnarray}
Thus, we can rewrite the spatial momentum integral as
\begin{eqnarray}
\int \frac{d^3k}{(2\pi)^3} &=& \int\limits_0^\infty \frac{k^2
  dk}{4\pi^2} \int\limits_{-1}^1 dc  \nonumber \\ &=& \frac{1}{4\pi^2
  p}\int\limits_0^\infty \omega_2~d\omega_2
\int\limits_{\omega_1^-}^{\omega_1^+} \omega_1~d\omega_1,
\end{eqnarray}
where
\begin{eqnarray}
\omega_1^{\pm} = \left[(p\pm k)^2+m_1^2\right]^{1/2}.
\end{eqnarray}
Using the above relations, we can write Eq.~(\ref{J1}) as
\begin{eqnarray}
J_1 &=& \frac{1}{16\pi^2p}\int\limits_0^\infty d\omega_2
\int\limits_{\omega_1^-}^{\omega_1^+}
d\omega_1\left(\frac{(\omega_1+\omega_2)^2}{2}-\frac{p^2}{2}-2m_f^2\right)\Bigl[
  1-n_F(\omega_1)-n_F(\omega_2)\Bigr ]
\delta(p_0-\omega_1-\omega_2)\nonumber\\ &=&
\frac{(P^2-4m_f^2)}{32\pi^2p} \int\limits_{\omega_2^-}^{\omega_2^+}
~d\omega_2~\Bigl[1-n_F(p_0-\omega_2)-n_F(\omega_2)\Bigr]\Theta(P^2-(m_1+m_2)^2),
\label{fd_wfa_j1}
\end{eqnarray}
and similarly for Eqs.~(\ref{J2}) and (\ref{J3}),
\begin{eqnarray}
J_2 &=& \frac{(P^2-4m_f^2)}{32\pi^2p}
\int\limits_{\omega_2^-}^{\omega_2^+}
~d\omega_2~(\omega_2^2-m_2^2)\Bigl[1-n_F(p_0-\omega_2)-n_F(\omega_2)\Bigr]\Theta(P^2-(m_1+m_2)^2),\label{fd_wfa_j2}\\ J_3
&=& \frac{1}{16\pi^2p} \int\limits_{\omega_2^-}^{\omega_2^+}
~d\omega_2~(\omega_2^2-m_2^2)\Bigl[1-n_F(p_0-\omega_2)-n_F(\omega_2)\Bigr]\Theta(P^2-(m_1+m_2)^2),\label{fd_wfa_j3}
\end{eqnarray}
where the limits
$\omega_2^\pm$ are obtained utilizing the argument of the Dirac
delta-function as
\begin{eqnarray}
\omega_2^\pm = \frac{p_0R \pm
  p\sqrt{R^2-4P^2m_2^2}}{2P^2}~;~~R=P^2+m_2^2-m_1^2.\nonumber
\end{eqnarray}
The final expression of the decay width is given by substituting
Eqs.~(\ref{fd_wfa_j1}), (\ref{fd_wfa_j2}) and (\ref{fd_wfa_j3}) back
in Eq.~(\ref{fd_wfa_final}).

\section*{Acknowledgments}

Work partially supported by research grants from Conselho Nacional de
Desenvolvimento Cient\'{\i}fico e Tecnol\'ogico (CNPq), under grants
304758/2017-5 (R.L.S.F) and 302545/2017-4 (R.O.R); Funda\c{c}\~ao
Carlos Chagas Filho de Amparo \`a Pesquisa do Estado do Rio de Janeiro
(FAPERJ), grant No.  E - 26/202.892/2017 (R.O.R) and CAPES (A.B). A.B
would also like to thank A Ayala, N Haque and M G Mustafa for helpful
discussions. 



\end{document}